# Improved Row-Column-Addressed Array Imaging by Leveraging Ghost Echoes

Chung-Shiang Mei
*Dept. of Electrical Engineering*
*National Tsing Hua University*
Hsinchu, Taiwan
s111061535@m111.nthu.edu.tw

Meng-Lin Li
*Dept. of Electrical Engineering*
*Inst. Of Photonic Technologies*
*Brain Research Center*
*National Tsing Hua University*
Hsinchu, Taiwan
mlli@ee.nthu.edu.tw

*Abstract*—Among various choices of 2-D ultrasound transducer arrays, the row-column-addressed (RCA) 2-D array has shown its promise for 3-D imaging. However, RCA suffers from notable edge effects and thus receives ghost echoes, which result in ghost artifacts showing in the volumetric image. In this research, rather than discarding these ghost echoes, we consider them as supportive signals and incorporate them into a post-filtering-based method to enhance the RCA imaging quality. Field II simulation and results of a single scatterer are presented in this work. The strongest ghost artifact is suppressed by 23 dB and the -6 dB lateral resolution is improved from 1.06 to 0.81 mm. The proposed method shows promising results in suppressing ghost artifacts and enhancing lateral resolution.

*Keywords—volumetric imaging, row-column-addressed array, edge effect, ghost artifacts, post-filtering*

## I. Introduction

3-D ultrasound imaging is an emerging topic in real-time medical imaging. To deal with volumetric data, a fully addressed 2-D matrix array is adopted, which is composed of $N \times N$ elements. With such a large number of electronic channels, the fabrication complexity can be demanding, e.g., wires in a cable to be connected to an ultrasonic imaging system is a challenge [1]–[4]. To reduce the number of channel counts for a 2-D array, a row-column-addressed (RCA) array is viewed as one of the possible options. In general, the RCA array is composed of two 1-D arrays, which are orthogonally crossed with each other as shown in Fig. 1. Under the design of the electrode-crossed structure, the number of channels is significantly reduced from $N \times N$ to $N + N$.

Unlike fully-addressed and sparse arrays, the shape of the element is rectangular rather than square element. Since the element of RCA is elongated, the assumption that considers the element as a point detector is incapable for RCA. In [3], [4], the work describes the shape of the emitted waveform and formulates the proper way to find the propagation path for applying a conventional delay-and-sum algorithm. However, the RCA suffers from well-known edge effects [2]–[4]. Due to the rectangular shape of the element, the responses for each element occur three times instead of once as in the case of a regular point detector. Considering both transmit element and receive element, nine echoes in total would be received, where the strongest response is called the main echo and others are called ghost echoes. Because of ghost echoes, purely applying delay-and-sum algorithm would result in an image with multiple ghost artifacts. To tackle the phenomenon, [3], [4] proposed a hardware apodization, which adds a roll-off region along an element, to physically get rid of the responses from the edges of the element and successfully suppress ghost artifacts. Nonetheless, producing such an array may be complicated and flexibility is sacrificed once the hardware is manufactured.

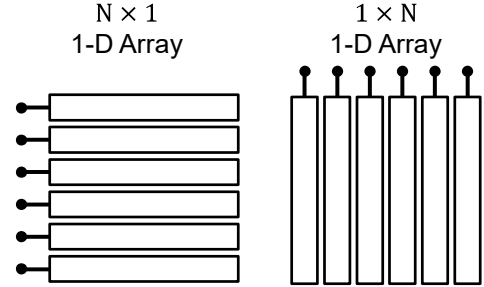

Fig. 1. RCA array is composed of two 1-D arrays.

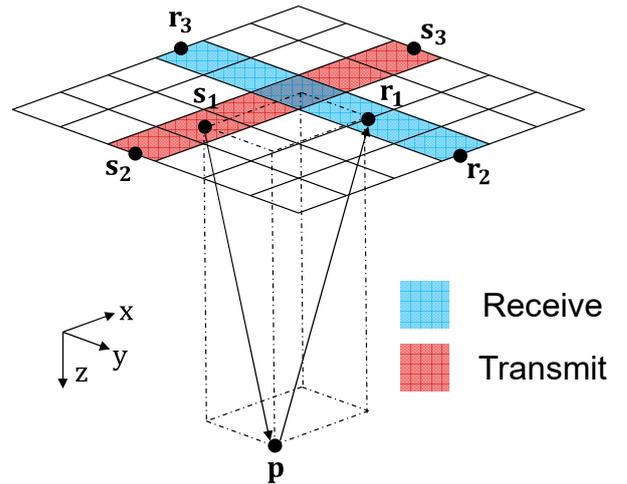

Fig. 2. Illustration of the paths of time-of-flight.

In this work, we propose a post-filtering-based method to suppress ghost artifacts. Other than discarding these ghost echoes, we leverage them as supportive signals. With these ghost echoes, we obtain eight more images by applying delay-and-sum beamforming according to the propagating path for each ghost echo. We call the image obtained from the main echo "main frame" and call the other eight images "ghost frames". Among these frames, we find the similarity and calculate the correlation factor. Once the correlation is evaluated, we utilize the correlation factor as a weighting map and multiply it onto the main frame. Ultimately, the main signal in the main frame is preserved and ghost artifacts are suppressed.

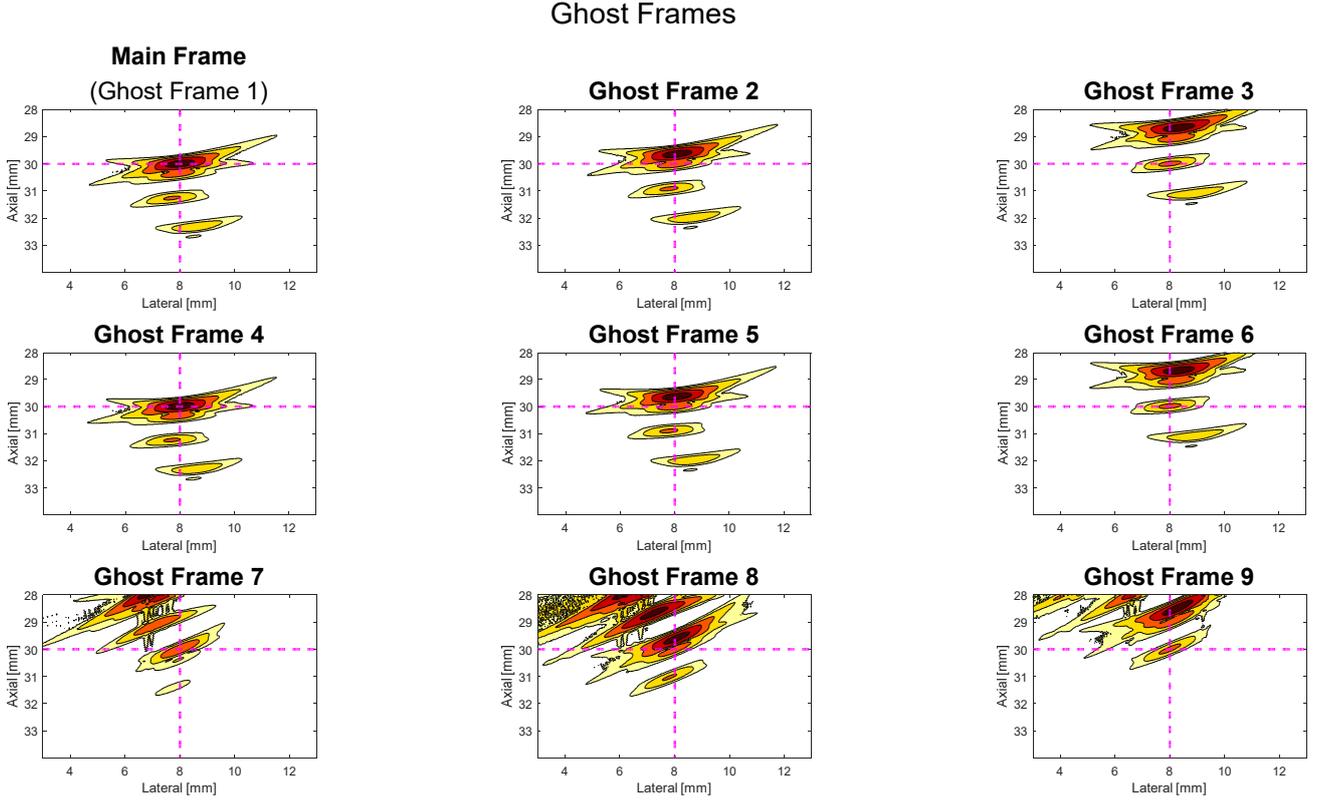

Fig. 3. Nine ghost frames obtained by applying delay-and-sum with different path for each echo.

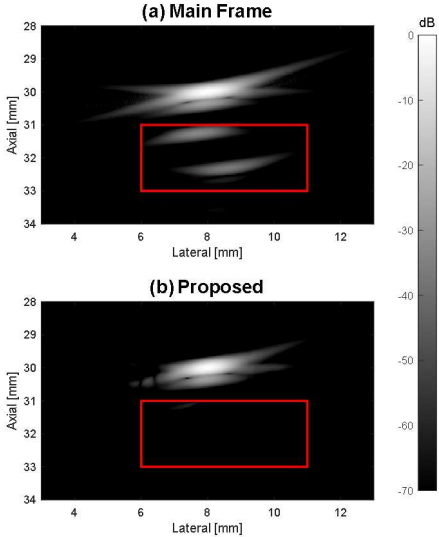

Fig. 4. Original main frame (a) and the image of PSF using the proposed method (b). Red boxes show the suppression of ghost artifacts.

## II. PROPOSED METHODS

### A. Ghost Frames

Considering the RCA echo propagation path, the time-of-flight (ToF) is described in [3], [4], which is

$$\text{ToF}(\mathbf{p}, n, i) = \frac{\|\mathbf{p}-\mathbf{s_n}\|+\|\mathbf{r_i}-\mathbf{p}\|}{c} \quad (1)$$

where n and i are both indices ranging from 1 to 3, $\mathbf{p}$ is the position of the scatterer, $\mathbf{s}$ is the position of transmit element, $\mathbf{r}$ is the position of receive element. The illustration is shown in Fig. 2 (see also in [3], [4]). In general, the path which considers the closest positions to the point $\mathbf{p}$, i.e., $\mathbf{s_1}$ and $\mathbf{r_1}$, are used to obtain the main frame. Paths which consider other than $\mathbf{s_1}$ or $\mathbf{r_1}$ are the positions of edges and hence those are used to obtain ghost frames. After applying delay-and-sum with nine different paths, nine images in total can be obtained, as shown in Fig. 3. Compared with the main frame, eight ghost frames are a rotated, transformed version of the main frame. In addition, the positions of ghost artifacts for each ghost frame are inconsistently located, but it is not the case for the position of the actual scatterer (intersection of the dashed line). Hence, the observation implies that we can detect the main target signal by finding the similarities among every frame.

### B. Post-Filtering

To evaluate the similarity, we calculate the normalized correlation factor [5] between the main frame and the ghost frame. The complex correlation is defined as

$$\text{Corr}(x, y) = \frac{\sum_{i=1}^{n} xy^*}{\sqrt{\sum_{i=1}^{n}\|x\|^2}\sqrt{\sum_{i=1}^{n}\|y\|^2}} \quad (2)$$

where x, y are baseband data of two frames, n is the number of voxels in the image. Based on the observation of ghost frames, the correlation factor map should be close to one if the position is where the actual scatterer is located and vice versa. After calculating the correlation map, we treat it as a weighting map and multiply it onto the main frame to get the result.

TABLE I.   SIMULATIONS PARAMETERS

| Parameter Name | Value | Unit |
|---|---|---|
| Center frequency | 5.0 | MHz |
| Speed of sound | 1480 | m/s |
| Wave length | 296 | μm |
| Array pitch (x) | 148 | μm |
| Array picth (y) | 148 | μm |
| Sampling frequency | 120 | MHz |
| Emission pulse | 2-cycles, Hann-weighted | - |

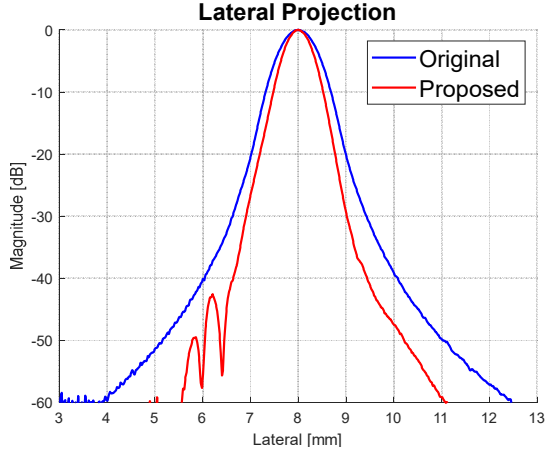

Fig. 5.  Lateral projection of the resulting PSF.

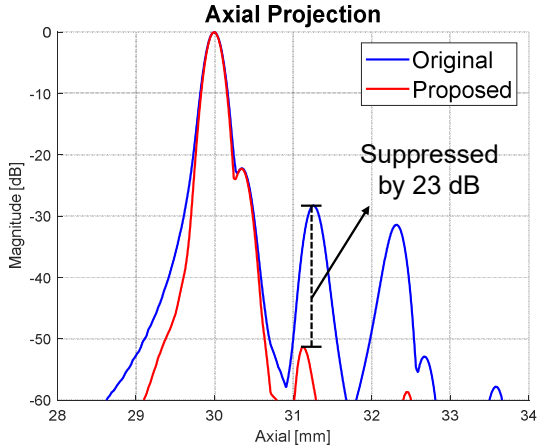

Fig. 6.  Axial projection of the resulting PSF.

## III. SIMULATION RESULTS

The simulation of a single scatterer located at (x, y, z) = (8, 3, 30) mm is presented and is done on Field II toolbox [6], [7] in this work. The simulation specification follows [3] and [4], which is used to verify the correctness of the original image with ghost artifacts. Details of simulation parameters are shown in Table I. The delay-and-sum beamforming is done using the synthetic aperture imaging [8]. Besides, Hanning apodization is adopted when receiving.

For the point spread function (PSF), the main frame shown in Fig. 4 (a) presents apparent ghost artifacts that appear in the image as shown in the red box. In contrast, ghost artifacts are suppressed in the resulting image obtained by the proposed method as shown in Fig. 4 (b).

In the comparison of projected profiles shown in Figs. 5 and 6, the -6 dB lateral resolution is improved from 1.06 to 0.81 mm. Also, the strongest ghost artifact is suppressed by 23 dB estimated from the axial projection shown in Fig. 6.

Other than the evaluation of PSF, an anechoic vessel is also simulated (not shown here) and the proposed method results in improved contrast and clearer vessel edge definition.

## IV. CONCLUSIONS

In this work, a post-filtering-based method that utilizes ghost echoes as supportive signals and focuses on ghost artifact suppression is proposed. Simulations of a single scatterer is presented to show the effectiveness of the method. The result shows that the ghost artifacts are greatly suppressed and the lateral resolution is improved. To sum up, the proposed method is capable of 3-D volumetric imaging using RCA.

ACKNOWLEDGMENTS

This work is supported by National Science and Technology Council, Taiwan under the grant number MOST 110-2221-E-007-011-MY3.

REFERENCES

[1] C. H. Seo and J. T. Yen, "A 256 x 256 2-D array transducer with row-column addressing for 3-D rectilinear imaging," *IEEE Transactions on Ultrasonics, Ferroelectrics, and Frequency Control*, vol. 56, no. 4, pp. 837-847, April 2009.

[2] M. F. Rasmussen and J. A. Jensen, "3D ultrasound imaging performance of a row-column addressed 2D array transducer: a simulation study ," *Proc. SPIE* Medical Imaging, March 2013.

[3] M. F. Rasmussen, T. L. Christiansen, E. V. Thomsen and J. A. Jensen, "3-D imaging using row-column-addressed arrays with integrated apodization - part i: apodization design and line element beamforming," *IEEE Transactions on Ultrasonics, Ferroelectrics, and Frequency Control*, vol. 62, no. 5, pp. 947-958, May 2015.

[4] T. L. Christiansen, M. F. Rasmussen, J. P. Bagge, L. N. Moesner, J. A. Jensen and E. V. Thomsen, "3-D imaging using row–column-addressed arrays with integrated apodization— part ii: transducer fabrication and experimental results," *IEEE Transactions on Ultrasonics, Ferroelectrics, and Frequency Control*, vol. 62, no. 5, pp. 959-971, May 2015.

[5] C. H. Seo and J. T. Yen, "Sidelobe suppression in ultrasound imaging using dual apodization with cross-correlation," *IEEE Transactions on Ultrasonics, Ferroelectrics, and Frequency Control*, vol. 55, no. 10, pp. 2198-2210, October 2008.

[6] J. A. Jensen and N. B. Svendsen, "Calculation of pressure fields from arbitrarily shaped, apodized, and excited ultrasound transducers," IEEE *Transactions on Ultrasonics, Ferroelectrics, and Frequency Control*, vol. 39, no. 2, pp. 262-267, March 1992.

[7] J. A. Jensen, "Field: A program for simulating ultrasound systems", *Med. Biol. Eng. Comput.*, vol. 34, no. 1, pp. 351-353, 1996.

[8] J. A. Jensen, S. I. Nikolov, K. L. Gammelmark and M. H. Pedersen, "Synthetic aperture ultrasound imaging," *Ultrasonics*, vol. 44, 2006.